# Effects of attachment preferences on coevolution of opinions and networks

Li-Xin Zhong,[1] Fei Ren,[2] Tian Qiu,[3] Jiang-Rong Xu,[1] Bi-Hui Chen[1]

[1]School of Journalism, Hangzhou Dianzi University, Hangzhou 310018, China

[2]School of Business, East China University of Science and Technology, Shanghai 200237, China

[3]School of Electronic and Information Engineering, Nanchang Hangkong University, Nanchang, 330063, China

In the coevolution of network structures and opinion formation, we investigate the effects of a mixed population with distinctive relinking preferences on both the convergence time and the network structures. It has been found that a heterogeneous network structure is easier to be reached with more high-degree-preferential(HDP) nodes. There exists high correlation between the convergence time and the network heterogeneity. The heterogeneous degree distribution caused by preferential attachment accelerates the convergence to a consensus state and the shortened convergence time inhibits the occurrence of the following disquieting situation that occurs in a continuously evolving network: with preferential attachment and long-time evolvement, most of the nodes would become separated and only a few leaders would have immediate neighbors. Analytical calculations based on mean field theory reveal that both the transition point $p_{tr}$ and the consensus time $t_c$ depend upon the standard deviation of the degree distribution $\sigma_d$. $p_{tr}$ increases while $t_c$ decreases with the rise of $\sigma_d$. Functions of $p_{tr} = \frac{<k>}{<k>+1}$ and $t_c = \frac{NK}{\sigma_d^2 - K}$ are found. Theoretical analyses are in accordance with simulation data.

## I. INTRODUCTION

In recent years, the study of collective behaviors in social and economic systems have attracted much interest among physicists [1-7]. Just like the disordered spin flip dynamics in the physical world, a seemingly simple competition between two contradictory opinions can lead to the occurrence of complexity in social dynamics, such as the nonlinear diffusion of public opinion and the dynamic fragmentation and recombination transitions in population structures [8].

There are various versions of opinion formation models, such as the voter model [9], the Sznajd model [10], the Deffuant-Weisbuch model [11] and the Hegselmann-Krause model [12], in which the discrete or continuous opinion dynamics have been discussed. In a binary-choice opinion formation model, each individual is endowed with a binary-state variable s=1 or s=-1 (spin up or spin down). By adopting a neighbor's contradictory opinion or convincing the neighbor to change his state, the population may finally converge to an absorbing consensus state or keep an active steady state with a constant fraction of competing opinions [13].

In real-life societies, as the agents with competing opinions meet, they may update their opinions or break the existed links and build up new links. The interplay between the states of interacting nodes and the physics of networks has recently been studied extensively [14-19]. One of the most important findings is that, in the coevolving opinion formation model, a slight modification in the updating rule can drastically change the system behavior[20]. Therefore, it is natural for us to ask such a question: if we have a modification in the reconnecting rule, can the model's behavior also be changed?

The people around us have their own preferences about when, where or how to cut the existed links and find new friends. There always exist some people who prefer linking to authority figures, sometimes called opinion leaders[21]. In a continuously growing social network, the existence of these people will produce "the rich get richer" effect[22]. But in a continuously evolving graph with a constant number of edges and vertices, such a preferential attachment will lead to the occurrence of a pessimistic scenario in which most of the nodes become separated from each other[23].

In real life, the nongrowing social network, the preferential attachment and the intimately connected people are observed widely. for example, the members of a party may rewire their mutual connections according to their own preferences but the party group as a whole still exists. To understand the real social dynamics, in the present work, we introduce a mixed population with high-degree-preferential (HDP) nodes, who tend to make an attachment to a node with high degree, and same-state-preferential (SSP) nodes, who prefer relinking to a node with the same opinion, into the opinion formation model introduced by F. Vazquez et al [13]. In social and economic fields, the HDP or SSP nodes correspond to trend followers or the individuals with independent personality separately. Following the work done in ref.[13], we have three main findings:

(1) The convergence time is highly related to the relinking preference of the population. In comparison with random attachment, HDP attachment not only promotes the formation of a connected population with the same opinion but also accelerates the convergence to a consensus state.

(2) There exists a transition point $p_{tr}$ of rewiring probability, below which the network forms a single component and above which the network becomes disconnected. More HDP nodes will have the the value of $p_{tr}$ increased. Such a result implies that, with more HDP nodes, the evolved network is not easy to be fragmented into disconnected components.

(3) With regard to the relationship between the convergence time $t_c$ and the standard deviation $\sigma_d$ of the degree distribution, we find a high correlation between the shortened convergence time and the limited network heterogeneity. With the increase of the number of HDP nodes, $t_c$ decreases with the rise of $\sigma_d$ which in turn weakens the heterogeneity of the evolving network. Therefore, the pessimistic scenario found in the continuously evolving nongrowing network model does not occur in the present model. As we focus on the change of active links between the nodes for the time being, we find that the strong correlation between $t_c$ and $\sigma_d$ is analytically manageable with mean field theory.

The plan of the paper is as follows. We introduce the coevolving opinion formation model with mixed population in Section II. In Section III, results of detailed simulations are presented and discussed. In Section IV, we give analytical calculations of $\sigma_d$, $t_c$ and $p_{tr}$ within the context of mean field theory. A summary is given in Section V.

## II. MODEL

Consider a network with N nodes, in which each node is initially connected to exactly the same number of immediate neighbors. There are two possible states +1 and −1 for each node, which may represent two contradictory opinions in human beings or up and down states in spin glasses. The states of the nodes and the links of the network evolve according to the following rules.

At each time step, a randomly chosen node i compares his opinion with his immediate neighbor j's. If they have the same opinion, nothing happens. Or else, node i rewires the link between them with probability p or imitates node j's state with probability 1-p. In the following, we call the links connecting nodes with different states active links and those between nodes in the same state inert link*s*.

To observe how the mixed population affects the coevolution of the strategies and the stuctures, and also to make the results analytically tractable, we divide the population into two categories: HDP nodes and SSP nodes. A parameter q is incorporated into the model to represent the ratio of HDP nodes in the whole population. Therefore, the number of SSP nodes should be N(1-q). In the rewiring process, an HDP node i firstly breaks one of his active links and then makes a new link to any other node m with probability $\Pi^{HDP} = k_m / \Sigma_n k_n$, in which $k_m$ is the degree of node m and $\Sigma_n k_n$ represents the total degree summed over all the nodes. But for an SSP node i, he will make the new link to a randomly chosen node m whose opinion is the same as his.

The convergence time for the system to reach an absorbing state is related to the density of active links. In the absorbing state, no active link exists between the nodes and the system should be in one of the following two states: the uniform state, where the network is a single component and all the nodes share the same opinion, and the disconnected state, where the network is fragmented into several components and the nodes in the same component share the same opinion. Although an absorbing state may not always be reached in an infinite system, it is more possible for a finite system to get into such a state because of fluctuations. As the coevolution only occurs when an active link is chosen, the density of active links represents not only the possibility of rewiring but also the possibility of opinion change.

In the coevolving process, the average number of links per node, also called average degree, is kept constant while the degree distribution evolves. The standard deviation $\sigma_d$ of the degree distribution is used to measure the dispersion of degrees. In the final steady state, the size of the first or the second largest component can tell us whether the network is highly connected or not. If the size of the first largest component is about N, it tells us that the system is in a uniform state. If the size of the first largest component is nearly the same as that of the second largest component, it tells us that two opposing groups occur in the system. Therefore, in the whole paper, except in special cases, the size of the first or the second largest component is used to measure whether or not there exist opposing groups in the system.

### III. RESULTS AND DISCUSSIONS

We have performed extensive numerical simulations to study the effects of the presence of such a mixed population on the changes of the convergence time and the evolved network structure in the coevolving opinion formation model. We typically consider systems of N=1000 and K=3. Starting with a random regular network in which all the nodes have the same degree 2K and an initial configuration of states in which half of the nodes are in +1 state, all the data are obtained after the system have been in the steady state.

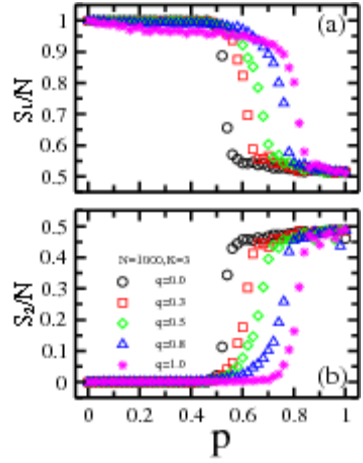

FIG.1. (Color online) Average relative size of (a) the largest network component and (b) the second largest network component in the stationary configuration as a function of rewiring probability p for N=1000, K=3, and q=0.0(circles), 0.3(squares), 0.5(diamonds), 0.8(triangles), 1.0(stars), averaged over 20 runs.

In Fig.1(a) and (b) we give the averaged relative size of the largest and the second largest network components as a function of the rewiring probability p with q=0, 0.3, 0.5, 0.8 and 1.0, when the system has already reached the absorbing state where only inert links exist. From Fig.1(a) we observe that, as p changes from 0 to unity, the relative size of the largest component in the final configuration changes from 1 to about 0.5. There exists a transition point $p_{tr}$, below which the network forms a single component and above which the network gets disconnected. In comparison with the results in Fig.1(b), we know that above $p_{tr}$ the network has fragmented into two large components accompanied with a set of small components of which the size is much smaller than N, indicating that two balanced opposition groups are easy to occur in the fast rewiring process. The transition point $p_{tr}$ increases monotonically with the rise of q, that is, the increase of HDP nodes in the population. Such a result indicate that, in the rewiring process, high degree preferential attachment is beneficial for the occurrence of a well-connected population with a uniform state.

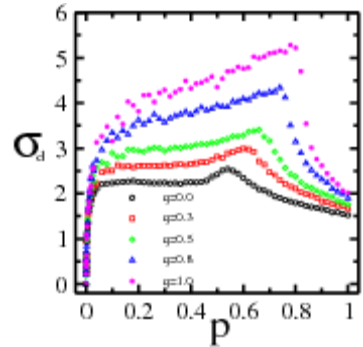

FIG.2. (Color online) The averaged standard deviation $\sigma_d$ of degree distribution as a function of rewiring probability p for N=1000, K=3, and different values of q, averaged over 20 runs

To get a close eye of the network stucture in the final frozen state, in Fig. 2 we plot the standard deviation $\sigma_d$ of the degree distribution as a function of p with q=0, 0.3, 0.5, 0.8 and 1.0. The results show that, for small and intermediate p, which corresponds to the uniform states in Fig.1(a) and (b), $\sigma_d$ increases with the rise of p and reaches a maximum at the transition point $p_{tr}$. For large p, which corresponds to disconnected states in Fig.1(a) and (b), $\sigma_d$ drops sharply with the rise of p. Note that within the whole range of $0<p\leqslant 1$, increasing q leads to an obvious increase of $\sigma_d$.

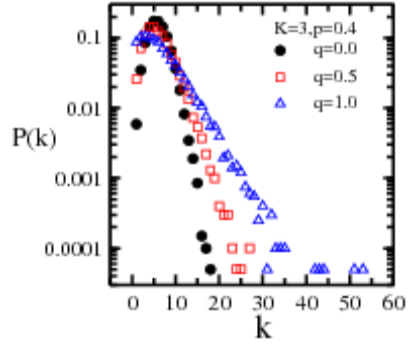

FIG.3. (Color online) The degree distribution of the evolved network for N=1000, K=3, p=0.4, and q=0(circles), 0.5(squares), 1.0(triangles), averaged over 20 runs.

In Fig.3 we give the degree distribution P(k) of the evolved network in the final frozen state with a fixed p and different values of q. From Fig. 3 we observe that, for q=0, P(k) shows a Possion distribution, just like that in a random network, and most of the nodes have the average degree 2K. As q increases, the nodes with higher degrees begin to occur and an exponential decay of P(k) is found for q=1. This finding suggests that, for a population with more HDP nodes, opinion learders are much easier to occur. It may be the existence of these opinion leaders that plays a dominant role in the changed system behavior.

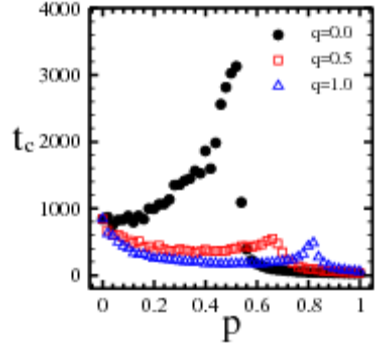

FIG.4. (Color online) Average convergence time $t_c$ as a function of rewiring probability p for N=1000, K=3, and q=0(circles), 0.5(squares), and 1.0(triangles). The averages are over 100 runs.

The average convergence time $t_c$ reflects the velocity for the system to reach an absorbing state, which is a magnitude of interest for scientists in knowing of how the random walk reaches unanimity. In Fig. 3 we give $t_c$ as a function of p for q=0, 0.5, and 1.0. For small p and q, $t_c$ increases sensitively with the rise of p. The value of $t_c$ reaches a maximum of $t_{max}$ at the critical point $p_{tr}$ which is also found in Fig.1 and Fig.2. For a large q, such an increase is not so distinct but the slowing down of the convergence to the final state is also found at the critical point. For $p>p_{tr}$, in all the curves $t_c$ decays sharply with the rise of p and the scaling of

$$t_c \sim \frac{\ln\left[2K(p-p_{tr})N\right]}{2K(p-p_{tr})}$$

, which has already been reported in ref.[13], is found. One of the novel findings in the present model is that, within the range of $0<p<p_{tr}$, $t_c$ decreases obviously with the rise of q, indicating that a global consensus is easier to be reached for the population with more HDP nodes.

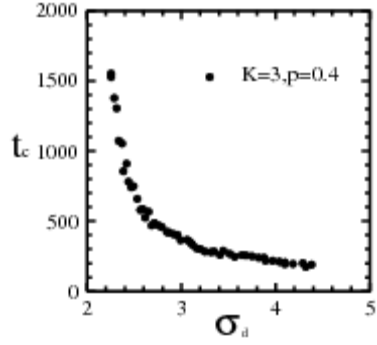

FIG.5. The average convergence time $t_c$ is plotted against $\sigma_d$ for p=0.4 and $0 \leqslant q \leqslant 1$. The results are obtained by averaging over 100 runs.

It is a special interest for us to investigate qualitative and quantitative relationships between the shortened convergence time and the evolved network structure. Comparing the simulation results in Fig. 4 with that in Fig. 2, we surprisingly find that, for a fixed p, increasing q leads to the rise of $\sigma_d$ but the drop of $t_c$. In Fig. 5 we give the convergence time $t_c$ as a function of $\sigma_d$ for p=0.4 and $0 \leqslant q \leqslant 1$. It is observed that the convergence time has an obvious dependency on the degree distribution of the evolved network. Within the whole range of $0 \leqslant q \leqslant 1$, $t_c$ decreases monotonously with the rise of $\sigma_d$. Similar dependence of $t_c$ on the mth moment of the degree distribution in static networks has been reported in ref.[9].

## IV. ANALYTICAL CALCULATIONS

Analytically, we can estimate the convergence time $t_c$ and the critical point $p_c$ by considering the degree distribution of the evolved network.

### A. relationship between the convergence time $t_c$ and the standard deviation $\sigma_d$ of the degree distribution

Consider a nongrowing network model in which the links between nodes can be removed and reestablished constantly in time. Recent studies have shown that, with random or preferential attachment, a final steady state can be reached and the evolved network has a Possion, an exponential or a power-law degree distribution [24].

Different from a continuously evolving network model, in the present model, the network structure only evolves when active links exist. From simulation results we find that, because of the limited evolving time, the evolved network shows an equilibrium configuration with both Possion and exponential degee distributions. The mechanism responsible for the occurrence of such a degree distribution should be related to the mixed population with distinctive attachment preferences.

From Fig.3 we observe that, as q increases, the degree distributuion of the network evolves from Possion to exponential distribution. For analytical calculations, the evolved network can be viewed as the combination of two uncorrelated networks: a random network with connectivity probability 1-q and average degree $<k_1>=2(1-q)\rho K$, and a growing network with an exponential degree distribution and average degree $<k_2>=2q\rho K$, where $\rho$ represents the average ratio of active links. For large N, the degree distribution of the random network is a

Possion distribution

$$P'(k_1) = \frac{[2(1-q)\rho K]^{k_1}}{k_1!} \exp[-2(1-q)\rho K]. \quad (1)$$

As for the degree distribution of the growing network, we adopt the exponential form

$$P''(k_2) = \frac{e}{q\rho K} \exp(-\frac{k_2}{q\rho K}), \quad (2)$$

which was found in the limiting case of the scale-free model [25].

The degree of a node can be written as $k = k_1 + k_2$ and the degree distribution P(k) of the evolved network is

$$P(k) = \sum_{k_2=0}^{k} \frac{[2(1-q)\rho K]^{k-k_2}}{(k-k_2)!} \exp[-2(1-q)\rho K] \frac{e}{q\rho K} \exp(-\frac{k_2}{q\rho K}). \quad (3)$$

Therefore, $\sigma_d$ can be written as

$$\sigma_d = \sqrt{4q^2 \rho^2 K^2 + 2(1-q)\rho K} \quad (4)$$

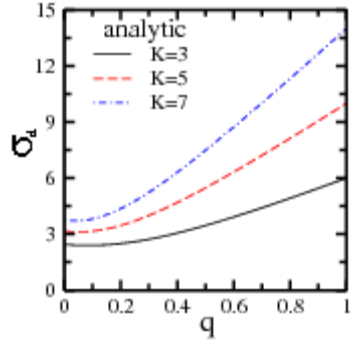

FIG.6. (Color online) The standard deviation $\sigma_d$ of the degree distribution predicted by Mean Field Theory.

In Fig.6 we plot the standard deviation $\sigma_d$ vs q according to equation (4) for different values of K=3, 5, 7, and $\rho = 1$. It was found in ref.[9] that, in a voter model, the acceleration of the convergence to a consensus state was related to the heterogeneity of the network. For a fixed K, the convergence time was determined by the total number of nodes in the network and the second moment of the degree distribution. That is,

$$t_c \sim \frac{N}{<k^2>}. \quad (5)$$

Examining the K-dependence of $t_c$ in the simulation results, we find the convergence time in the present model is also related to the average degree of the evolved network. Therefore, we perform a fit to the results in Fig. 5 to the following functional form

$$t_c = \frac{NK}{\sigma_d^2 - K}, \quad (6)$$

that is,

$$t_c = \frac{NK}{4q^2K^2 + 2(1-q)K - K} \quad (7)$$

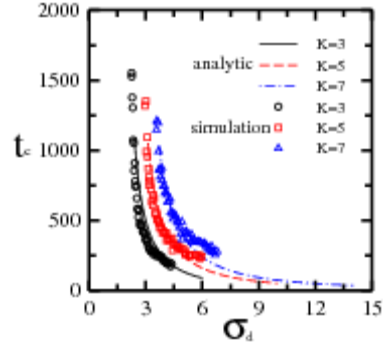

FIG.7. (Color online) The average convergence time $t_c$ vs $\sigma_d$ for K=3, 5, and 7 (dashed lines) predicted by Mean Field Theory.

In Fig.7 we plot $t_c$ as a function of $\sigma_d$ for K=3, 5 and 7. The dashed lines represent the theoretical predictions and the circles, squares, and triangles represent the simulation data. Our results show that, in the coevolution of network structures and opinion formation, the convergence time is dominated by the standard deviation $\sigma_d$ of the degree distribution. This finding suggests that it is the existence of nodes with higher degrees due to preferential relinking that plays a dominant role in determining the convergence time.

### B. relationship between the transition point $p_{tr}$ and the ratio q of population

In the coevolutionary dynamics, a phase transition in the network structure is often observed. In the present model, there exists a transition point $p_{tr}$ of the rewiring probability, below which the evolved nework is a single connected network with all the nodes adopting the same strategy and above which the evolved nework is fragmented into two or more disconnected components with the nodes in the same component adopting the same strategy.

In Ref.[13], the authors show that the average density $\rho$ of active links changes according to the following master equation

$$\frac{d\rho}{dt} = \sum_k P_k \frac{2}{k<k>}[(1-p)(k<n>_k - 2<n^2>_k) - p<n>_k], \quad (8)$$

in which $P_k$ is the node degree distribution at time $t$, $<n> = \sum_{n=0}^{k} nB_{n,k}$ and $<n^2> = \sum_{n=0}^{k} n^2 B_{n,k}$.

$B_{n,k}$ is the probability that n active links are connected to a node of degree k. In the mean field spirit, we suppose that the probability that a given link is active by the average density $\rho$. Therefore, $B_{n,k}$ should have the same form as P(k). For the mixed degree distribution

$$P(k) = \sum_{k_2=0}^{k} \frac{[2(1-q)\rho K]^{k-k_2}}{(k-k_2)!} \exp[-2(1-q)\rho K] \frac{e}{q\rho K} \exp(-\frac{k_2}{q\rho K}), \quad (9)$$

we get

$$<n>_k = \rho k \quad (10)$$

and

$$<n^2>_k = (1+q^2)\rho^2 k^2 + (1-q)\rho k. \quad (11)$$

Therefore, the master equation (8) becomes

$$\frac{d\rho}{dt} = \frac{2\rho}{<k>}[(1-p)(<k> -2q^2\rho<k> -2(1-q) -2\rho<k>) - p]. \quad (12)$$

For q=0, we get the ratio of active links

$$\rho_s = \frac{<k>(1-p) - 2(1-q)(1-p) - p}{2<k>(q^2+1)(1-p)} \quad (13)$$

and the transition point

$$p_{tr} = \frac{<k>-2}{<k>-1}. \quad (14)$$

For q=1, the ratio of active links becomes

$$\rho_s' = \frac{<k>(1-p) - p}{4<k>(1-p)} \quad (15)$$

and the transition point

$$p_{tr}' = \frac{<k>}{<k>+1}. \quad (16)$$

For a fixed $<k>$, it is observed that

$$p_{tr} < p_{tr}'. \quad (17)$$

Analytical calculations show that, as the degree distribution changes from a Possion to an exponential distribution, the transition point $p_{tr}$ increases accordingly. Such a result is in agreement with simulation results found in Fig.2. That is, the transition point increases with the rise of q.

## V. SUMMARY

We have studied the effects of a mixed population with distinctive relinking preferences on the opinion and community formation. It is observed that, with the increase of the number of HDP nodes in the population, the degree distribution of the network gradually evolves from a Possion through an exponential to a limited power-law degree distribution, and the consensus time drops accordingly. Analytical calculations show that the rapid drop of the consensus time is related to the heterogeneity of the evolved network, which in turn inhibits the occurrence of such a disquieting situation where nearly all the nodes are separated from each other. Therefore, depending upon the present model, we can understand why the population are still intimately connected to each other in real society although it continuously evolves.

In the future, the mechanism responsible for the coevolution of network structures and collective behaviors should be further studied in the models competing for limited resources in social or natural society, such as that in the evolutionary minority game or the snowdrift game, which may help us obtain an insight into the continuously evolving process in real society.

## ACKNOWLEDGMENTS

This work is the research fruit of Zhejiang Social Sciences Association under Grant 08N51,

"Chen Guang" project supported by Shanghai Municipal Education Commission and Shanghai Education Development Foundation (2008CG37), and the National Natural Science Foundation of China (Grant Nos. 10905023, 10805025, and 10774080).